\documentclass[aip,jcp,reprint,graphicx]{revtex4-1}

\draft % marks overfull lines with a black rule on the right

\usepackage{graphicx}
\usepackage{booktabs}
\usepackage{multirow}
\usepackage{amsmath}
\usepackage{amsfonts}
\usepackage{color}
\usepackage[caption=false]{subfig}

\graphicspath{{./figures/}}

\newcommand{\R}{\mathbf{R}}
\newcommand{\rr}{\mathbf{r}}
\newcommand{\x}{\mathbf{x}}
\newcommand{\ba}{\mathbf{a}}

\newcommand{\new}[1]{#1}

\begin{document}

%\preprint{}

\title{SchNet -- a deep learning architecture for molecules and materials} %Title of paper



\author{K.T. Sch\"utt}
\email[]{kristof.schuett@tu-berlin.de}

\affiliation{Machine Learning Group, Technische Universit\"at Berlin, 10587 Berlin, Germany}

\author{H.E. Sauceda}
\affiliation{Fritz-Haber-Institut der Max-Planck-Gesellschaft, 14195 Berlin, Germany}

\author{P.-J. Kindermans}
\affiliation{Machine Learning Group, Technische Universit\"at Berlin, 10587 Berlin, Germany}

\author{A. Tkatchenko}
\email[]{alexandre.tkatchenko@uni.lu}
\affiliation{Physics and Materials Science Research Unit, University of Luxembourg, L-1511 Luxembourg, Luxembourg}

\author{K.-R. M\"uller}
\email[]{klaus-robert.mueller@tu-berlin.de}
\affiliation{Machine Learning Group, Technische Universit\"at Berlin, 10587 Berlin, Germany}
\affiliation{Max-Planck-Institut f\"ur Informatik, Saarbr\"ucken, Germany}
\affiliation{Department of Brain and Cognitive Engineering, Korea University, Anam-dong, Seongbuk-gu,
Seoul 136-713, South Korea}

\date{\today}

\begin{abstract}
Deep learning has led to a paradigm shift in artificial intelligence, including web, text and image search, speech recognition, as well as bioinformatics, with growing impact in chemical physics. Machine learning in general and deep learning in particular is ideally suited for representing quantum-mechanical interactions, enabling to model nonlinear potential-energy surfaces or enhancing the exploration of chemical compound space.
Here we present the deep learning architecture SchNet that is specifically designed to model atomistic systems by making use of continuous-filter convolutional layers.
We demonstrate the capabilities of SchNet by accurately predicting a range of properties across chemical space for \emph{molecules and materials} where our model learns chemically plausible embeddings of atom types across the periodic table.
Finally, we employ SchNet to predict potential-energy surfaces and energy-conserving force fields for molecular dynamics simulations of small molecules and perform an exemplary study of the quantum-mechanical properties of C$_{20}$-fullerene that would have been infeasible with regular \textit{ab initio} molecular dynamics.
\end{abstract}


\maketitle %\maketitle must follow title, authors, abstract and \pacs

\section{Introduction}
\label{sec:introduction}

% If in two-column mode, this environment will change to single-column format so that long equations can be displayed. 
%\begin{widetext}
%$$\mbox{put long equation here}$$
%\end{widetext}

% Figures should be put into the text as floats. 
% Use the graphics or graphicx packages (distributed with LaTeX2e).
% Put the label that you will use with \ref{} command in the braces of the \label{} command.
%
% \begin{figure}
% \end{figure}

% \begin{table}
% \end{tabular}
% \end{table}
Accelerating the discovery of molecules and materials with desired properties is a long-standing challenge in computational chemistry and the materials sciences.
However, the computational cost of accurate quantum-chemical calculations proves prohibitive in the exploration of the vast chemical space.
In recent years, there have been increased efforts to overcome this bottleneck using machine learning, where only a reduced set of reference calculations is required to accurately predict chemical properties~\citep{rupp2012fast,montavon2013machine,hansen2013assessment,schutt2014represent,faber2015crystal,ramakrishnan2015big,Hansen-JCPL,faber2016machine,hirn2017wavelet,faber2017fast,huo2017unified,eickenberg2017scattering,isayev2017universal,ryczko2017convolutional,luchak2017extensive} or potential-energy surfaces~\citep{behler2007generalized,behler2011atom,bartok2010gaussian,bartok2013representing,shapeev2016moment,chmiela2017machine,brockherde2017bypassing,smith2017ani,podryabinkin2017active,rowe2017machine}.
While these approaches make use of painstakingly handcrafted descriptors, deep learning has been applied to predict properties from molecular structures using graph neural networks~\citep{duvenaud2015convolutional,Kearnes2016}.
However, these are restricted to predictions for equilibrium structures due to the lack of atomic positions in the input.
Only recently, approaches that learn a representation directly from atom types and positions have been developed~\citep{schutt2017quantum, gilmer2017neural, schutt2017schnet}.
While neural networks are often considered a 'black box', there has recently been an increased effort to explain their predictions in order to understand how they operate or even extract scientific insight.
This can either be done by analyzing a trained model~\citep{Baehrens2010,simonyan2013deep,bach2015pixel,Zintgraf2017,Montavon2017,kindermans2017learning,montavon2018methods} or by directly designing interpretable models~\citep{xu2015show}.
For quantum chemistry, some of us have proposed such an interpretable architecture with Deep Tensor Neural Networks (DTNN) that not only learns a representation of atomic environments but allows for spatially and chemically resolved insights into quantum-mechanical observables~\citep{schutt2017quantum}.

Here we build upon this work and present the deep learning architecture SchNet that allows to model complex atomic interactions in order to predict potential-energy surfaces or speeding up the exploration of chemical space.
SchNet, being a variant of DTNNs, is able to learn representations for molecules and materials that follow fundamental symmetries of atomistic systems by construction, e.g., rotational and translational invariance as well as invariance to atom indexing.
This enables accurate predictions throughout compositional and configurational chemical space where symmetries of the potential energy surface are captured by design.
Interactions between atoms are modeled using continuous-filter convolutional layers~\citep{schutt2017schnet} being able to incorporate further chemical knowledge and constraints using specifically designed filter-generating neural networks.
We demonstrate that those allow to efficiently incorporate periodic boundary conditions enabling accurate predictions of formation energies for a diverse set of bulk crystals.
Beyond that, both SchNet and DTNNs provide local chemical potentials to analyze the obtained representation and allow for chemical insights~\citep{schutt2017quantum}.
An analysis of the obtained representation shows that SchNet learns chemically plausible embeddings of atom types that capture the structure of the periodic table.
Finally, we present a path-integral molecular dynamics (PIMD) simulation using an energy-conserving force field learned by SchNet trained on reference data from a classical MD at the PBE+vdW$^{\rm TS}$~\cite{PBE1996,TS2009} level of theory effectively accelerating the simulation by three orders of magnitude. Specifically, we employ the recently developed perturbed path-integral approach~\cite{PPI2016} for carrying out imaginary time PIMD, which allows quick convergence of quantum-mechanical properties with respect to the number of classical replicas (beads).
This exemplary study shows the advantages of developing computationally efficient force fields with \textit{ab initio} accuracy, allowing nanoseconds of PIMD simulations at low temperatures -- an inconceivable task for regular \textit{ab initio} molecular dynamics (AIMD) that could be completed with SchNet within hours instead of years.

\section{Method}
\label{sec:theory}

\begin{figure}
	\includegraphics[width=0.95\columnwidth]{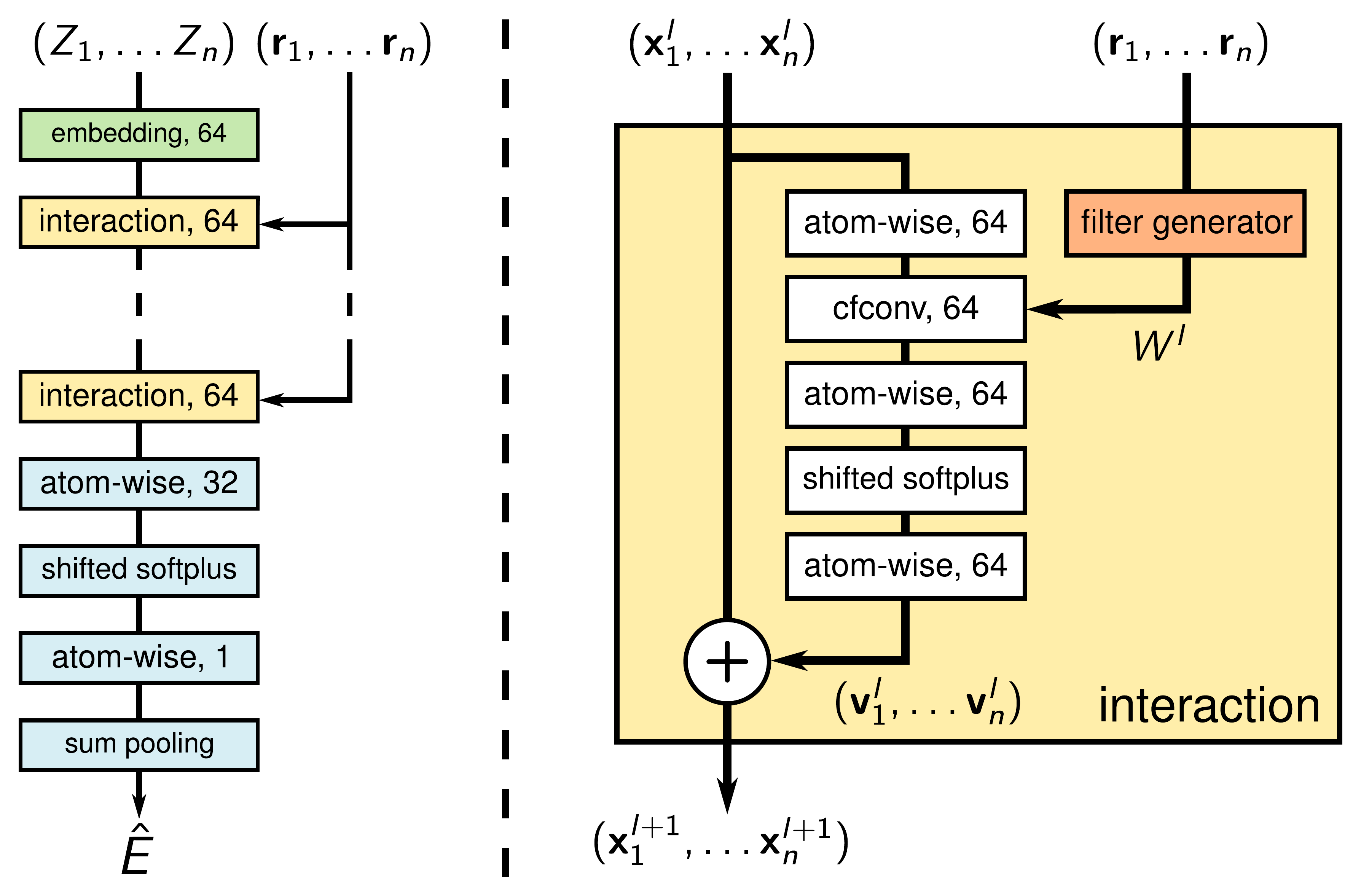}
	\caption{Illustrations of the SchNet architecture (left) and interaction blocks (right) with atom embedding in green, interaction blocks in yellow and property prediction network in blue. \new{For each parameterized layer, the number of neurons is given.} The filter-generating network (orange) is shown in detail in Fig. \ref{fig:filter}.}\label{fig:overview}
\end{figure}

SchNet is a variant of the earlier proposed Deep Tensor Neural Networks (DTNN)~\citep{schutt2017quantum} and therefore shares a number of their essential building blocks.
Among these are atom embeddings, interaction refinements and atom-wise energy contributions.
At each layer, the atomistic system is represented atom-wise being refined using pair-wise interactions with the surrounding atoms.
\new{In the DTNN framework, interactions are modeled by tensor layers, i.e., atom representations and interatomic distances are combined using a parameter tensor. This can be approximated using a low-rank factorization for computational efficiency~\citep{Taylor2009a,yu2013deep,socher2013recursive}.
SchNet instead makes use of continuous-filter convolutions with filter-generating networks~\citep{BrabandereJTG16,schutt2017schnet} to model the interaction term.}
These can be interpreted as a special case of such factorized tensor layers.
In the following, we introduce these components and describe how they are assembled to form the SchNet architecture.
For an overview of the SchNet architecture, see Fig.~\ref{fig:overview}.

\subsection{Atom embeddings}

An atomistic system can be described uniquely by a set of $n$ atom sites with nuclear charges $Z=(Z_1, \dots, Z_n)$ and positions $R=(\rr_1, \dots \rr_n)$.
Through the layers of SchNet, the atoms are described by a tuple of features $X^l= (\x_1^l, \dots \x_n^l)$, with $\x^l_i \in \mathbb{R}^F$ with the number of feature maps $F$, the number of atoms $n$ and the current layer $l$.
The representation of site $i$ is initialized using an embedding dependent on the atom type $Z_i$:
\begin{equation}
\x^0_i = \ba_{Z_i}.
\end{equation}
These embeddings $\ba_Z$ are initialized randomly and optimized during training.
They represent atoms of a system disregarding any information about their environment for now.

\subsection{Atom-wise layers}
Atom-wise layers are dense layers that are applied separately to the representations $\x^{l}_i$ of each atom $i$:
\begin{equation}
\x^{l+1}_i = W^l \x^{l}_i + \mathbf{b}^l
\end{equation}
Since weights $W^l$ and biases $\mathbf{b}^l$ are shared across atoms, our architecture remains scalable with respect to the number of atoms.
While the atom representations are passed through the network, these layers transform them and process information about the atomic environments incorporated through interaction layers.

\subsection{Interaction blocks}

The interaction blocks of SchNet add refinements to the atom representation based on pair-wise interactions with the surrounding atoms.
In contrast to DTNNs, here we model these with continuous-filter convolutional layers (cfconv) that are a generalization of the discrete convolutional layers commonly used, e.g., for images~\citep{lecun1989backpropagation, krizhevsky2012imagenet} or audio data~\citep{van2016wavenet}.
This generalization is necessary since atoms are not located on a regular grid like image pixels, but can be located at arbitrary positions.
\new{Therefore, a filter-tensor, as used in conventional convolutional layers, is not applicable. Instead we need to model the filters continuously with a filter-generating neural network.}
Given atom-wise representations $X^l$ at positions $R$, we obtain the interactions of atom $i$ as the convolution with all surrounding atoms
\begin{equation}
\x^{l+1}_i = (X^l * W^l)_i = \sum_{j=0}^{n_\text{atoms}} \x_j^l \circ W^l(\rr_j - \rr_i),
\end{equation}
where "$\circ$" represents the element-wise multiplication. 
Note that we perform feature-wise convolutions for computational efficiency~\citep{chollet2016xception}.
Cross-feature processing is subsequently performed by atom-wise layers.
Instead of a filter tensor, we define a filter-generating network $W^l: \mathbb{R}^3 \rightarrow \mathbb{R}^F$ that maps the atom positions to the corresponding values of the filter bank (see Section \ref{sec:fgn}).

A cfconv layer together with three atom-wise layers constitutes the residual mapping~\citep{he2016deep} of an interaction block (see Fig.~\ref{fig:overview}, right).
We use a shifted softplus $\text{ssp}(x) = \ln(0.5 e^x + 0.5)$ as activation functions throughout the network.
The shifting ensures that $\text{ssp}(0) = 0$ and improves the convergence of the network \new{while having infinite order of continuity.
This allows us to obtain smooth potential energy surfaces, force fields and second derivatives that are required for training with forces as well as the calculation of vibrational modes.}

\subsection{Filter-generating networks}\label{sec:fgn}
\begin{figure}
	\includegraphics[width=\columnwidth]{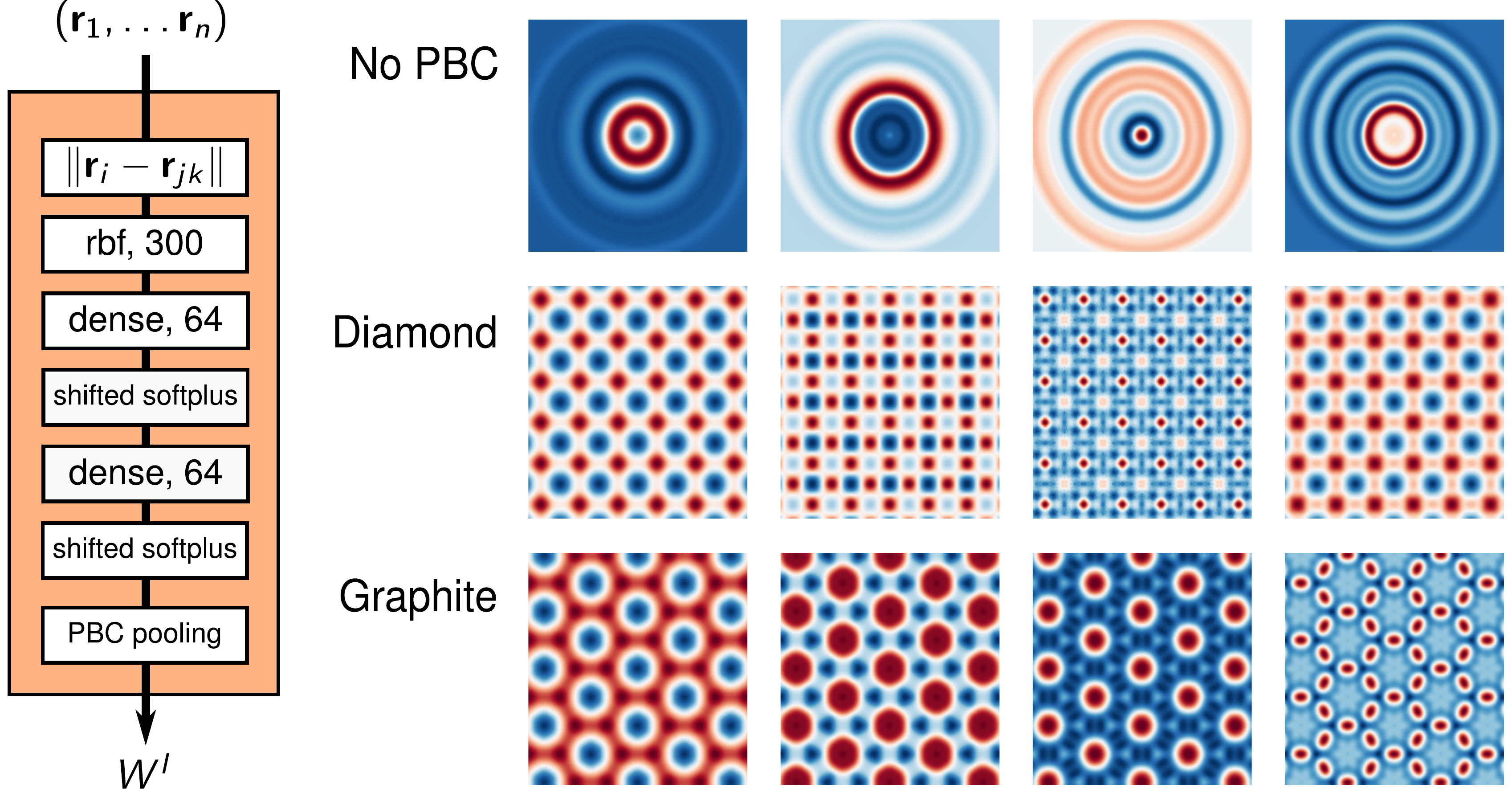}
	\caption{Architecture of the filter-generating network used in SchNet (left) and 5{\AA} x 5{\AA} cuts through generated filters (right) from the same filter-generating networks (columns) under different periodic bounding conditions (rows). Each filter is learned from data and represents the effect of an interaction on a given feature of an atom representation located in the center of the filter. \new{For each parameterized layer, the number of neurons is given.}}\label{fig:filter}
\end{figure}

The filter-generating network determines how interactions between atoms are modeled and can be used to constrain the model and include chemical knowledge.
We choose a fully-connected neural network that takes the vector pointing from atom $i$ to its neighbor $j$ as input to obtain the filter values $W(\rr_j - \rr_i)$ (see Fig.~\ref{fig:filter}, left).
This allows us to include known invariances of molecules and materials into the model.

\subsubsection{Rotational invariance}
It is straightforward to include rotational invariance by computing pairwise distances instead of using relative positions.
We further expand the distances in a basis of Gaussians
\[
e_k(\rr_j-\rr_i) = \exp ( -\gamma (\|\rr_j-\rr_i\| - \mu_k )^2 ),
\]
with centers $\mu_k$ chosen on a uniform grid between zero and the distance cutoff.
This has the effect of decorrelating the filter values which improves the conditioning of the optimization problem.
The number of Gaussians and the hyper parameter $\gamma$ determine the resolution of the filter.
\new{We have set the grid spacing and scaling parameter $\gamma$ to be $0.1${\AA} for all models in this work.}

\subsubsection{Periodic boundary conditions}
For atomistic systems with periodic boundary conditions (PBCs), each atom-wise feature vector $\x_i$ has to be equivalent across all periodic repetitions, i.e., $\x_i = \x_{ia} = \x_{ib}$ for repeated unit cells $a$ and $b$.
Due to the linearity of the convolution, we are therefore able to apply the PBCs directly to the filter to accurately describe the atom interactions while keeping invariance to the choice of the unit cell.
Given a filter $\tilde{W}^l(\rr_{jb}-\rr_{ia})$ over all atoms with $\|\rr_{jb}-\rr_{ia}\| < r_{\text{cut}}$, we obtain the convolution
\begin{align*}
\x^{l+1}_{i} = \x^{l+1}_{im} &= \frac{1}{n_\text{neighbors}} \sum_{\substack{j, n \\ \rr_{jn}}} \x_{jn}^l \circ \tilde{W}^l(\rr_{jn} - \rr_{im}) \\
&= \frac{1}{n_\text{neighbors}} \sum_{j} \x^l_j \circ \underbrace{\left( \sum_n \tilde{W}^l(\rr_{jn} - \rr_{im}) \right)}_{W}.
\end{align*}
This new filter $W$ now depends on the PBCs of the system as we sum over all periodic images within the given cutoff $r_{\text{cut}}$.
We find that the training is more stable when normalizing the filter response $\x^{l+1}_{i}$ by the number of atoms within the cutoff range.
Fig.~\ref{fig:filter} (right) shows a selection of generated filters without PBCs, with a cubic diamond crystal cell and with an hexagonal graphite cell.
As the filters for diamond and graphite are superpositions of single-atom filters according to their respective lattice, they reflect the structure of the lattice.
Note that while the single-atom filters are circular due to the rotational invariance, the periodic filters become rotationally equivariant w.r.t. the orientation of the lattice, which still keeps the property prediction rotationally invariant.
While we have followed a data-driven approach where we only incorporate basic invariances in the filters, careful design of the filter-generating network provides the possibility to incorporate further chemical knowledge in the network.

\subsection{Property prediction}
Finally, a given property $P$ of a molecule or material is predicted from the obtained atom-wise representations.
We compute atom-wise contributions $\hat{P}_i$ from the fully-connected prediction network (see blue layers in Fig~\ref{fig:overview}).
Depending on whether the property is intensive or extensive, we calculate the final prediction $\hat{P}$ by summing or averaging over the atomic contributions, respectively.

\new{Since the initial atom embeddings are obviously equivariant to the order of atoms, atom-wise layers are independently applied to each atom and continuous-filter convolutions sum over all neighboring atoms, indexing equivariance is retained in the atom-wise representations.
Therefore, the prediction of properties as a sum over atom-wise contributions guarantees indexing invariance.}

When predicting atomic forces, we instead differentiate a SchNet predicting the energy w.r.t. the atomic positions:
\begin{multline}
\hat{\textbf{F}}_i(Z_1, \dots, Z_n, \rr_1, \dots, \rr_n) = \\
 -\frac{\partial \hat{E}}{\partial \rr_i}(Z_1, \dots, Z_n, \rr_1, \dots, \rr_n).
\end{multline}
When using a rotationally invariant energy model, this ensures rotationally equivariant force predictions and guarantees an energy conserving force field~\cite{chmiela2017machine}.

\subsection{Training}
We train SchNet for each property target $P$ by minimizing the squared loss
\[
\ell(\hat{P}, P) = \|P - \hat{P} \|^2.
\]
For the training of energies and forces of molecular dynamics trajectories, we use a combined loss
\begin{multline}
\ell((\hat{E}, \mathbf{\hat{F}}_1, \dots, \mathbf{\hat{F}}_n)), (E, \mathbf{F}_1, \dots, \mathbf{F}_n)) = \\
\rho \quad \|E - \hat{E} \|^2 + \frac{1}{n_\text{atoms}} \sum_{i=0}^{n_\text{atoms}} \left \| \mathbf{F}_i - \left (-\frac{\partial \hat{E}}{\partial \R_i}\right ) \right\|^2 \label{eq:loss}
\end{multline}
where $\rho$ is a trade-off between energy and force loss~\citep{pukrittayakamee2009simultaneous}.

All models are trained with mini-batch stochastic gradient descent using the ADAM optimizer~\citep{KingmaB14} with mini-batches of 32 examples.
We decay the learning rate exponentially with ratio $0.96$ every 100,000 steps.
In each experiment, we split the data into a training set of given size $N$ and use a validation set for early stopping. The remaining data is used for computing the test errors.
Since there is a maximum number of atoms being located within a given cutoff, the computational cost of a training step scales linearly with the system size if we precompute the indices of nearby atoms.

\section{Results}
\label{sec:results}

%We have applied SchNet to various settings ranging from property prediction across chemical compound space over formation energies of a diverse set of bulk crystals to potential energy surfaces and force fields of molecular dynamics simulations.
%As an application of our method, we have conducted a path integral molecular dynamics simulation of the fullerene C$_{20}$.

\subsection{Learning molecular properties}
%\subsubsection{QM9}

\begin{table}
\caption{\label{tab:qm9} Mean absolute errors for energy predictions on the QM9 data set using 110k training examples. For SchNet, we give the average over three repetitions as well as standard errors \new{of the mean of the repetitions}. Best models in bold.}
\begin{ruledtabular}
\begin{tabular}{llrr}
	Property               & Unit     &                      SchNet ($T=6$) & enn-s2s~\citep{gilmer2017neural} \\ \hline
	$\epsilon_\text{HOMO}$ & eV       &  \textbf{0.041 $\pm$ 0.001} &                            0.043 \\
	$\epsilon_\text{LUMO}$ & eV       &  \textbf{0.034 $\pm$ 0.000} &                            0.037 \\
	$\Delta\epsilon$       & eV       &  \textbf{0.063 $\pm$ 0.000} &                            0.069 \\
    ZPVE                   & meV      &             1.7 $\pm$ 0.033 &                     \textbf{1.5} \\
	$\mu$                  & Debye    &           0.033 $\pm$ 0.001 &                   \textbf{0.030} \\
	$\alpha$               & Bohr$^3$ &           0.235 $\pm$ 0.061 &                   \textbf{0.092} \\
	$\langle R^2 \rangle$  & Bohr$^2$ &  \textbf{0.073 $\pm$ 0.002} &                            0.180 \\
	$U_0$                  & eV       &  \textbf{0.014 $\pm$ 0.001} &                            0.019 \\
	$U$                    & eV       &  \textbf{0.019 $\pm$ 0.006} &                   \textbf{0.019} \\
	$H$                    & eV       &  \textbf{0.014 $\pm$ 0.001} &                            0.017 \\
	$G$                    & eV       &  \textbf{0.014 $\pm$ 0.000} &                            0.019 \\
	$C_v$                  & cal / molK & \textbf{0.033 $\pm$ 0.000} &                             0.040
\end{tabular}
\end{ruledtabular}
\end{table}

\begin{figure}
	\includegraphics[width=\columnwidth]{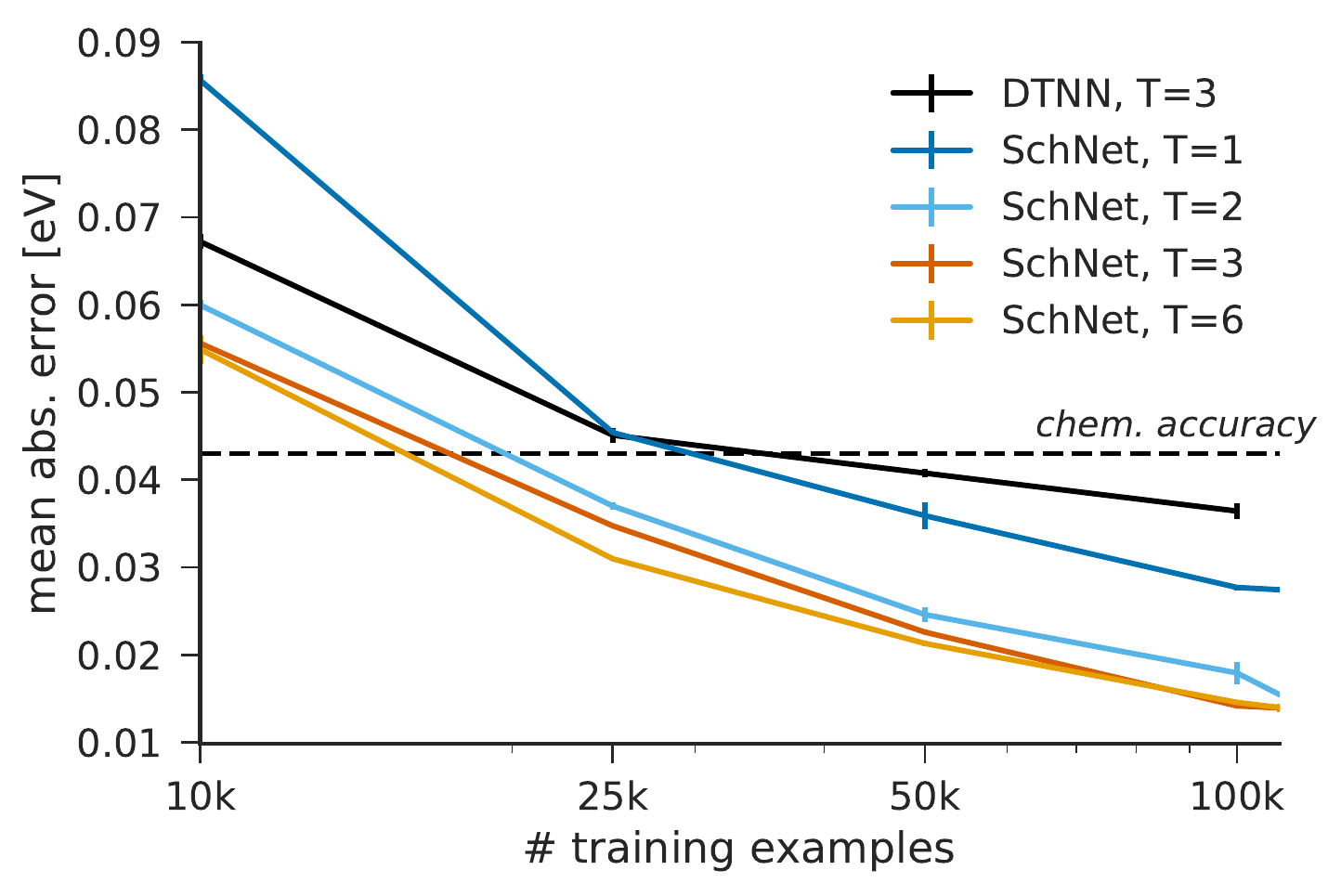}
	\caption{\label{fig:lc_qm9_U0} Mean absolute error (in eV) of energy predictions ($U_0$) on the QM9 dataset~\cite{ramakrishnan2014quantum,blum2009gdb13,reymond2015chemical} depending on the number of interaction blocks and reference calculations used for training. For reference, we give the best performing DTNN models \new{(T=3)}~\citep{schutt2017quantum}.}
\end{figure}

We train SchNet models to predict various properties of the QM9 dataset~\cite{ramakrishnan2014quantum,blum2009gdb13,reymond2015chemical} of 131k small organic molecules with up to nine heavy atoms from CONF.
Following~\citet{gilmer2017neural} and \citet{faber2017fast}, we use a validation set of 10,000 molecules.
We sum over atomic contribution $\hat{P}_i$ for all properties but $\epsilon_\text{HOMO}$, $\epsilon_\text{LUMO}$ and the gap $\Delta\epsilon$, where we take the average.
We use $T=6$ interaction blocks and atomic representations with $F=64$ feature dimension and perform up to 10 million gradient descent parameter updates.
\new{Since the molecules of QM9 are quite small, we do not use a distance cutoff. For the Gaussian expansion, we use a range up to 20{\AA} to cover all interatomic distances occurring in the data.}
The prediction errors are listed in Table~\ref{tab:qm9}, where we compare the performance to the message-passing neural network \emph{enn-s2s}~\citep{gilmer2017neural} that use additional bond information beyond atomic positions to learn a molecular representation.
The SchNet predictions of the polarizability $\alpha$ and the electronic spatial extent $\langle R^2 \rangle$ fall noticeably short in terms of accuracy.
This is most likely due to the decomposition of the energy into atomic contributions which is not appropriate for these properties. 
In contrast to SchNet, \citet{gilmer2017neural} employ a set2set model variant~\citep{vinyals2015order} that obtains a global representation and does not suffer from this issue.
However, SchNet reaches or improves over enn-s2s in 8 out of 12 properties where a decomposition into atomic contributions is a good choice.
\new{The distributions of the errors of all predicted properties are shown in Appendix~\ref{app:distribution}.}
Extending SchNet with interpretable, property-specific output layers, e.g. for the dipole moment~\citep{gastegger2017machine}, is subject to future work.

Fig.~\ref{fig:lc_qm9_U0} shows learning curves of SchNet for the total energy $U_0$ with $T \in \{1, 2, 3, 6\}$ interaction blocks compared to the best performing DTNN models~\citep{schutt2017quantum}.
\new{The best performing DTNN with $T=3$ interaction blocks can only outperform the SchNet model with $T=1$.}
We observe that beyond two interaction blocks the error improves only slightly from 0.015 eV with $T=2$ interaction blocks to 0.014 eV for $T \in \{3, 6\}$ using 110k training examples.
When training on fewer examples, the differences become more significant and $T=6$, while having the most parameters, exhibits the lowest errors.
Additionally, the model requires much less epochs to converge, e.g., using 110k training examples reducing the required number of epochs from $2400$ with $T=2$ to less than $750$ epochs with $T=6$.

%\subsubsection{Materials Project}

%\begin{table}
%	\caption{\label{tab:qm9} Mean absolute errors for formation energy predictions on the various materials data sets in meV/atom. For SchNet, we give the average over three repetitions as well as standard errors. Best models in bold. \note{preliminary}}
%	\begin{ruledtabular}
%		\begin{tabular}{lrrr}
%\textbf{Dataset}  & \textbf{SchNet}  & \textbf{\citet{faber2016machine}} & \textbf{MBTR~\citep{huo2017unified}} \\ \hline
%\textit{Materials Project} \\
%\quad N = 60k  & \textbf{35.2 $\pm$ 0.2} & --  & -- \\
%\textit{Elpasolites}  \\
%\, N = 9k, III-VI &  & -- & \textbf{2.7} \\
%\, N = 9k, I-VIII & \textbf{54.7} & 100.00 & --
%		\end{tabular}
%	\end{ruledtabular}
%\end{table}

\subsection{Learning formation energies of materials}

\begin{table}
	\caption{\label{tab:mp} Mean absolute errors for formation energy predictions in eV/atom on the Materials Project data set. For SchNet, we give the average over three repetitions as well as standard errors \new{of the mean of the repetitions}. Best models in bold.}
	\begin{ruledtabular}
		\begin{tabular}{lrr}
			Model              &  $N = 3,000$ & $N=60,000$ \\ \hline
     		\textbf{ext. Coulomb matrix}~\citep{faber2015crystal} & 0.64 & -- \\
			\textbf{Ewald sum matrix}~\citep{faber2015crystal} & 0.49 & -- \\
			\textbf{sine matrix}~\citep{faber2015crystal} & 0.37 & -- \\
			\textbf{SchNet} ($T=6$)   & \textbf{0.127 $\pm$ 0.001} & \textbf{0.035 $\pm$ 0.000}\\
		\end{tabular}
	\end{ruledtabular}
\end{table}

We employ SchNet to predict formation energies for bulk crystals using 69,640 structures and reference calculations from the Materials Project (MP) repository~\cite{Jain2013,Ong2012b}.
It consists of a large variety of bulk crystals with atom type ranging  across the whole periodic table up to $Z=94$.
Mean absolute errors are listed in Table \ref{tab:mp}.
Again, we use $T=6$ interaction blocks and atomic representations with $F=64$ feature dimension.
We set the distance cutoff $r_\text{cut}=5${\AA} and discard two examples from the data set that would include isolated atoms with this setting.
Then, the data is randomly split into 60,000 training examples, a validation set of 4,500 examples and the remaining data as test set.
Even though the MP repository is much more diverse than the QM9 molecule benchmark, SchNet is able to predict formation energies up to a mean absolute error of 0.035 eV/atom.
\new{The distribution of the errors is shown in Appendix~\ref{app:distribution}.}
On a smaller subset 3,000 training examples, SchNet still achieves an MAE of 0.127 eV/atom improving significantly upon the descriptors proposed by \citet{faber2015crystal}.

\begin{figure}
	\includegraphics[width=\columnwidth]{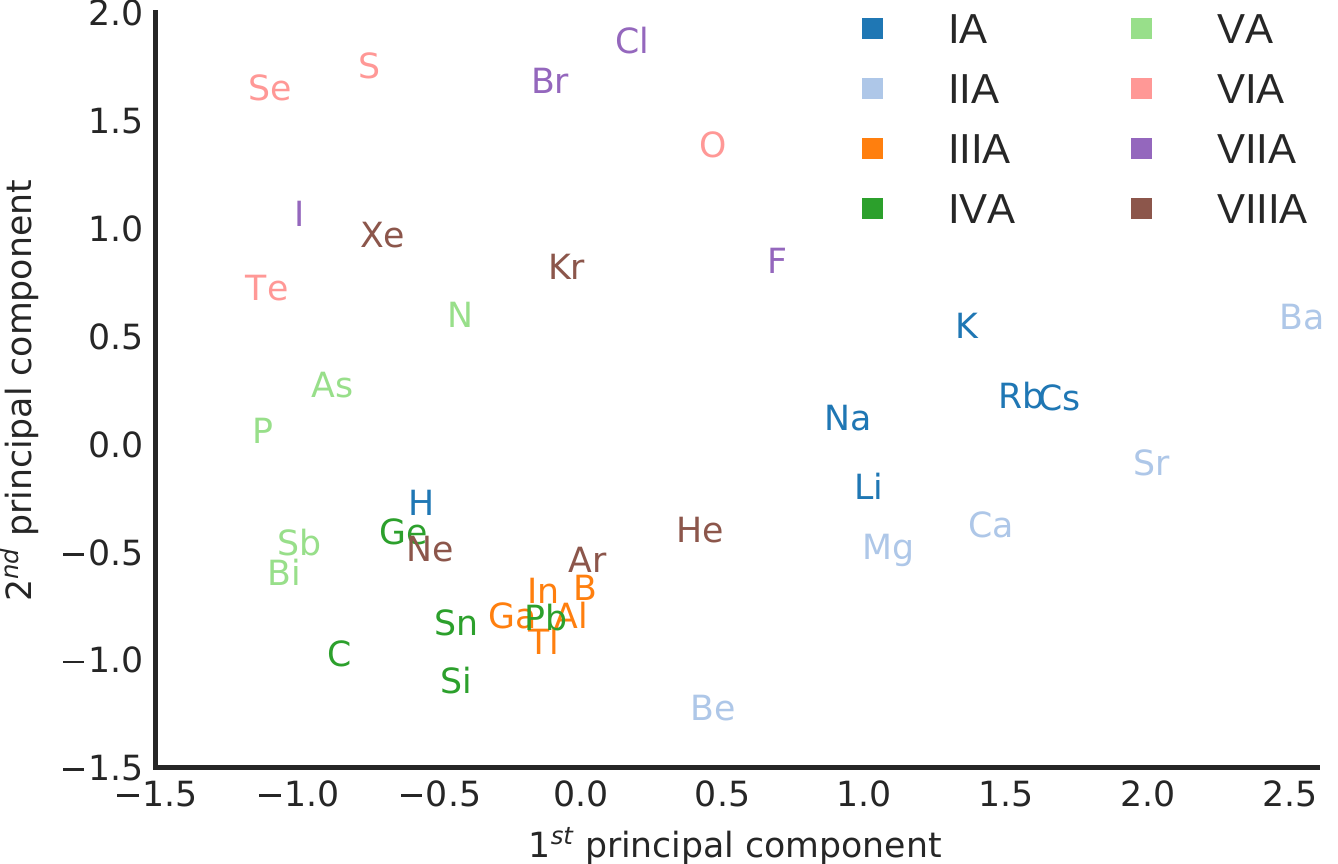}
	\caption{\label{fig:x0_mp} The two leading principal components of the learned embeddings $\mathbf{x}^0$ of sp atoms learned by SchNet from the Materials Project dataset. We recognize a structure in the embedding space according to the groups of the periodic table (color-coded) as well as an ordering from lighter to heavier elements within the groups, e.g., in groups IA and IIA from light atoms (left) to heavier atoms (right).}
\end{figure}

Since the MP dataset contains 89 atom types ranging across the periodic table, we examine the learned atom type embeddings $\mathbf{x}^0$.
Due to their high dimensionality, we visualize two leading principal components of all sp-atom type embeddings as well as their corresponding group (see Fig.~\ref{fig:x0_mp}).
The neural network aims to use the embedding space efficiently, such that this 2d projection explains only about 20\% of the variance of the embeddings, i.e., since important directions are missing, embeddings might cover each other in the projection while actually being further apart.
Still, we already recognize a grouping of elements following the groups of the periodic table.
This implies that SchNet has learned that atom types of the same group exhibit similar chemical properties.
Within some of the groups, we can even observe an ordering from lighter to heavier elements, e.g., in groups IA and IIA from light elements on the left to heavier ones on the right or, less clear in group VA with a partial ordering N -- \{As, P\} -- \{Sb, Bi\}.
Note that this knowledge was not imposed on the machine learning model, but inferred by SchNet from the geometries and formation energy targets of the MP data.

\subsection{Local chemical potentials}

\begin{figure*}
	\includegraphics[width=\textwidth]{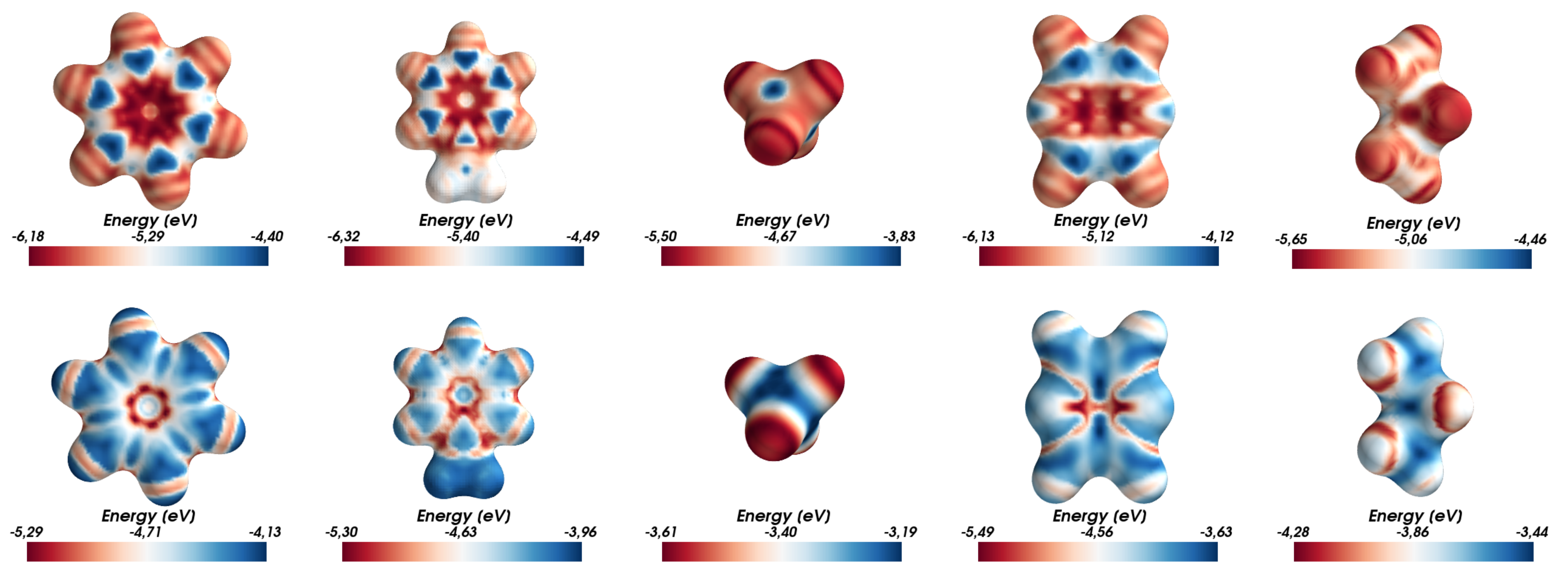}
	\caption{\label{fig:gbplots} Local chemical potentials $\Omega_C(\rr)$ of DTNN (top) and SchNet (bottom) using a carbon test charge on a $\sum_i \| \rr - \rr_i \|=3.7${\AA} isosurface are shown for benzene, toluene, methane, pyrazine and propane.}
\end{figure*}

\begin{figure}
	\includegraphics[width=\columnwidth]{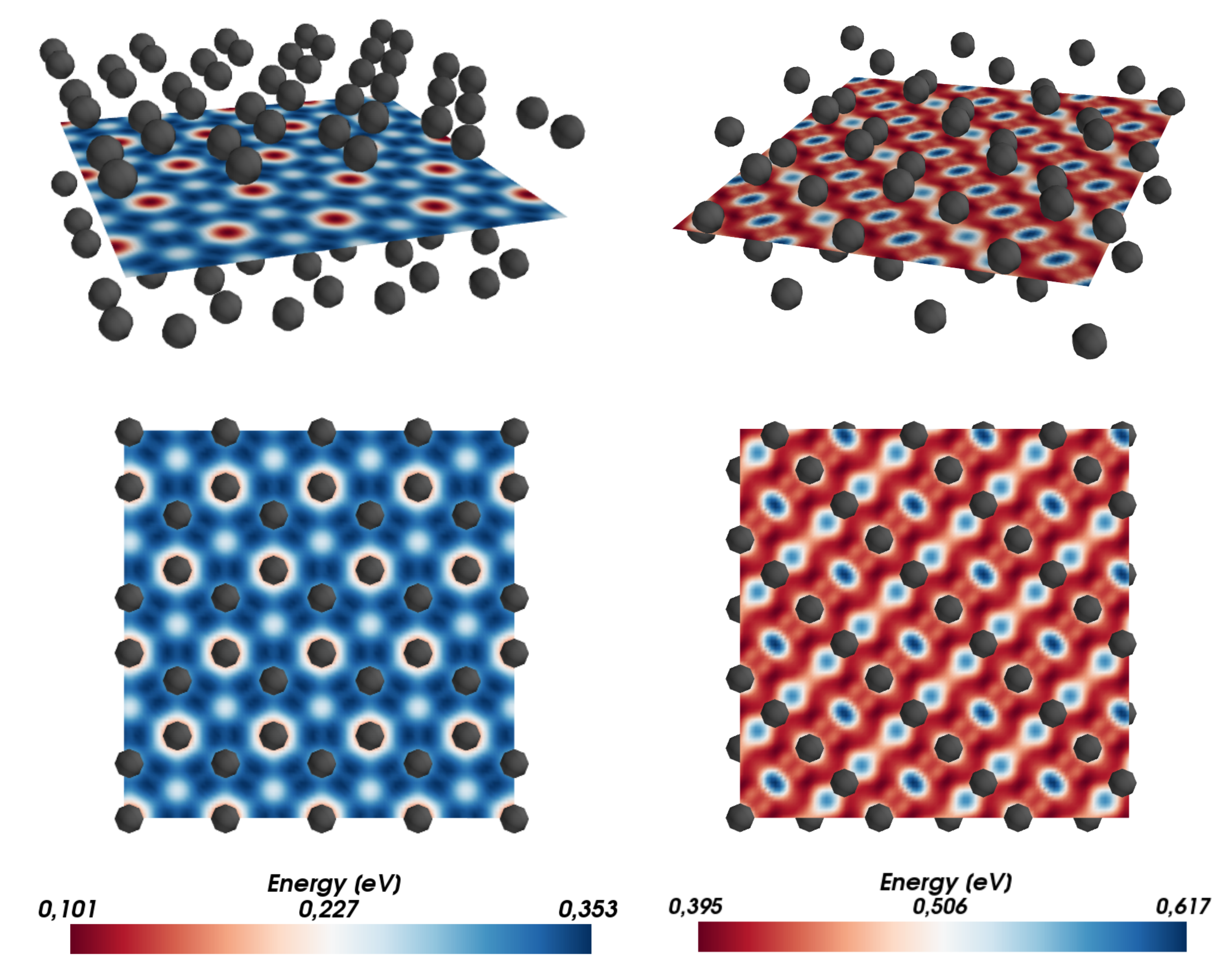}
	\caption{\label{fig:gbplotsmat} Cuts through local chemical potentials $\Omega_C(\rr)$ of SchNet using a carbon test charge are shown for graphite (left) and diamond (right).}
\end{figure}

Since the SchNet is a variant of DTNNs, we can visualize the learned representation with a ``local chemical potential'' $\Omega_{Z_\text{probe}}(\mathbf{r})$ as proposed by~\citet{schutt2017quantum}:
We compute the energy of a virtual atom that acts as a test charge.
This can be achieved by adding the probe atom $(Z_\text{probe}, \mathbf{r}_\text{probe})$ as an input of SchNet.
The continuous filter-convolution of the probe atom with the atoms of the system
\begin{equation}
\x^{l+1}_\text{probe} = (X^l * W^l)_i = \sum_{i=0}^{n_\text{atoms}} \x_i^l \circ W^l(\rr_\text{probe} - \rr_i),
\end{equation}
ensures that the test charge only senses but does not influence the feature representation.
We use Mayavi~\cite{ramachandran2011mayavi} to visualize the potentials.

Figure~\ref{fig:gbplots} shows a comparison of the local potentials of various molecules from QM9 generated by DTNN and SchNet. Both DTNN and SchNet can clearly grasp fundamental chemical concepts such as bond saturation and different degrees of aromaticity. While the general structure of the potential on the surfaces is similar, the SchNet potentials exhibit sharper features and have a more pronounced separation of high-energy and low-energy areas. The overall appearence of the distinguishing molecular features in the ``local chemical potentials'' is remarkably robust to the underlying neural network architecture, representing the common quantum-mechanical atomic embedding in its molecular environment. It remains to be seen how the ``local chemical potentials'' inferred by the networks can be correlated with traditional quantum-mechanical observables such as electron density, electrostatic potentials, or electronic orbitals. In addition, such local potentials could aid in the understanding and prediction of chemical reactivity trends.  

In the same manner, we show cuts through $\Omega_C(\rr)$ for graphite and diamond in Fig.~\ref{fig:gbplotsmat}.
As expected, they resemble the periodic structure of the solid, much like the corresponding filters in Fig.~\ref{fig:filter}. In solids, such local chemical potentials could be used to understand the formation and distribution of defects, such as vacancies and interstitials.

\subsection{Combined learning of energies and atomic forces}

%\subsubsection{MD17}
\begin{table*}
	\caption{\label{tab:mdenergies}Mean absolute errors for total energies (in kcal/mol). GDML~\citep{chmiela2017machine}, DTNN~\citep{schutt2017quantum} and SchNet~\citep{schutt2017schnet} test errors for  N=1,000 and N=50,000 reference calculations of molecular dynamics simulations of small, organic molecules are shown. Best results are given in bold.}
	\begin{ruledtabular}
		\begin{tabular}{lrrrrrr}
			                                 &                       \multicolumn{3}{c}{$N$ = 1,000}                        &                       \multicolumn{3}{c}{$N$ = 50,000}                       \\ \cline{2-4}\cline{5-7}
			                                 &   \textbf{GDML} &            \multicolumn{2}{c}{\textbf{SchNet}}             &   \textbf{DTNN} &            \multicolumn{2}{c}{\textbf{SchNet}}             \\ \cline{2-2}\cline{3-4}\cline{5-5}\cline{6-7}
			\raggedleft{\textit{trained on}} & \textit{forces} & \textit{energy} & \textit{energy+forces}  & \textit{energy} & \textit{energy} & \textit{energy+forces} \\ \hline
			Benzene                          &   \textbf{0.07} &            1.19 &                   0.08                  &   \textbf{0.04} &            0.08 &                   0.07  \\
			Toluene                          &   \textbf{0.12} &            2.95 &          \textbf{0.12}                  &            0.18 &            0.16 &          \textbf{0.09}  \\
			Malonaldehyde                    &            0.16 &            2.03 &         \textbf{ 0.13}                  &            0.19 &            0.13 &          \textbf{0.08}  \\
			Salicylic acid                   &   \textbf{0.12} &            3.27 &                   0.20                  &            0.41 &            0.25 &          \textbf{0.10}  \\
			Aspirin                          &   \textbf{0.27} &            4.20 &                   0.37                  &              -- &            0.25 &          \textbf{0.12}  \\
			Ethanol                          &            0.15 &            0.93 &          \textbf{0.08}                  &              -- &            0.07 &          \textbf{0.05}  \\
			Uracil                           &   \textbf{0.11} &            2.26 &                   0.14                  &              -- &            0.13 &          \textbf{0.10}  \\
			Naphthalene                       &  \textbf{ 0.12} &            3.58 &                   0.16                  &              -- &            0.20 &          \textbf{0.11} 
		\end{tabular}
	\end{ruledtabular}
\end{table*}

\begin{table*}
	\caption{\label{tab:mdforces}Mean absolute errors for atomic forces (in kcal/mol/\AA). GDML~\citep{chmiela2017machine} and SchNet~\citep{schutt2017schnet} test errors for N=1,000 and N=50,000 reference calculations of molecular dynamics simulations of small, organic molecules are shown. Best results are given in bold.}
	\begin{ruledtabular}
		\begin{tabular}{lrrrrr}
			                                 &                       \multicolumn{3}{c}{$N$ = 1,000}                        &              \multicolumn{2}{c}{$N$ = 50,000}              \\ \cline{2-4}\cline{5-6}
			                                 &   \textbf{GDML} &            \multicolumn{2}{c}{\textbf{SchNet}}             &            \multicolumn{2}{c}{\textbf{SchNet}}             \\ \cline{2-2}\cline{3-4} \cline{5-6}
			\raggedleft{\textit{trained on}} & \textit{forces} & \textit{energy} & \textit{energy+forces} &  \textit{energy} & \textit{energy+forces}  \\ \hline
			Benzene                          &   \textbf{0.23} &           14.12 &                   0.31 &                       1.23 &          \textbf{0.17}  \\
			Toluene                          &   \textbf{0.24} &           22.31 &                   0.57 &                          1.79 &          \textbf{0.09}   \\
			Malonaldehyde                    &            0.80 &           20.41 &          \textbf{0.66} &                           1.51 &          \textbf{0.08}   \\
			Salicylic acid                   &   \textbf{0.28} &           23.21 &                   0.85 &                           3.72 &          \textbf{0.19}   \\
			Aspirin                          &   \textbf{0.99} &           23.54 &                   1.35 &                           7.36 &          \textbf{0.33}   \\
			Ethanol                          &            0.79 &            6.56 &          \textbf{0.39} &                           0.76 &          \textbf{0.05}   \\
			Uracil                           &   \textbf{0.24} &           20.08 &                   0.56 &                           3.28 &          \textbf{0.11}   \\
			Naphthalene                       &   \textbf{0.23} &           25.36 &                   0.58 &                           2.58 &          \textbf{0.11} 
		\end{tabular}
	\end{ruledtabular}
\end{table*}

We apply SchNet to the prediction of potential energy surfaces and force fields of the MD17 benchmark set of molecular dynamics trajectories introduced by \citet{chmiela2017machine}.
MD17 is a collection of eight molecular dynamics simulations for small organic molecules.
Tables \ref{tab:mdenergies} and \ref{tab:mdforces} list mean absolute errors for energy and force predictions.
We trained SchNet on randomly sampled training sets with $N=1,000$ and $N=50,000$ reference calculations for up to 2 million mini-batch gradient steps and additionally used a validation set of 1,000 examples for early stopping.
The remaining data was used for testing.
We also list the performances of \new{gradient domain machine learning (GDML)}~\citep{chmiela2017machine} and DTNN~\citep{schutt2017quantum} for reference.
SchNet was trained with $T=3$ interaction blocks \new{and $F=64$ feature maps} using only energies as well as using the combined loss for energies and forces from Eq.~\ref{eq:loss} with $\rho=0.01$.
This trade-off constitutes a compromise to obtain a single model that performs well on energies and forces for a fair comparison with GDML.
\new{Again, we do not use a distance cutoff due to the small molecules and a range up to 20{\AA} for the Gaussian expansion to cover all distances.}
In Section~\ref{sec:c20}, we will see that even lower errors can be achieved when using two separate SchNet models for energies and forces.

SchNet can take significant advantage of the additional force information, reducing energy and force errors by 1-2 orders of magnitude compared to energy only training on the small training set.
With 50,000 training examples, the improvements are less apparent as the potential energy surface is already well-sampled at this point.
On the small training set, SchNet outperforms GDML on the more flexible molecules malonaldehyde and ethanol, while GDML reaches much lower force errors on the remaining MD trajectories that all include aromatic rings.
A possible reason is that GDML defines an order of atoms in the molecule, while the SchNet architecture is inherently invariant to indexing which constitutes a greater advantage in the more flexible molecules.

While GDML is more data-efficient than a neural network, SchNet is scalable to larger data sets.
We obtain MAEs of energy and force predictions below 0.12 kcal/mol and 0.33 kcal/mol/{\AA}, respectively.
Remarkably, SchNet performs better while using the combined loss with energies and forces on 1,000 reference calculations than training on energies of 50,000 examples.

\subsection{Application to molecular dynamics of C$_{20}$-fullerene}\label{sec:c20}

\begin{table}
	\caption{\label{tab:c20models} Mean absolute errors for energy and force predictions of C$_{20}$-fullerene in kcal/mol and kcal/mol/\AA, respectively. We compare SchNet models with varying number of interaction blocks $T$, feature dimensions $F$ and energy-force tradeoff $\rho$. For force-only training ($\rho=0$), the integration constant is fitted separately. Best models in bold.}
	\begin{ruledtabular}
		\begin{tabular}{lllrr}
			$T$ & $F$ & $\rho$ &          energy &          forces \\ \hline
			3   & 64  & 0.010  &          0.228 &          0.401 \\
			6   & 64  & 0.010  &          0.202 &          0.217 \\
			3   & 128 & 0.010  &          0.188 &          0.197 \\
			6   & 128 & 0.010  &          \textbf{0.1002} &         \textbf{ 0.120} \\ \hline
			6   & 128 & 0.100  & \textbf{0.027} &          0.171 \\
			6   & 128 & 0.010  &          0.100 &          0.120 \\
			6   & 128 & 0.001  &          0.238 &          0.061 \\
			6   & 128 & 0.000  &          0.260 & \textbf{0.058}
		\end{tabular}
	\end{ruledtabular}
\end{table}

\begin{figure}
	\includegraphics[width=\columnwidth]{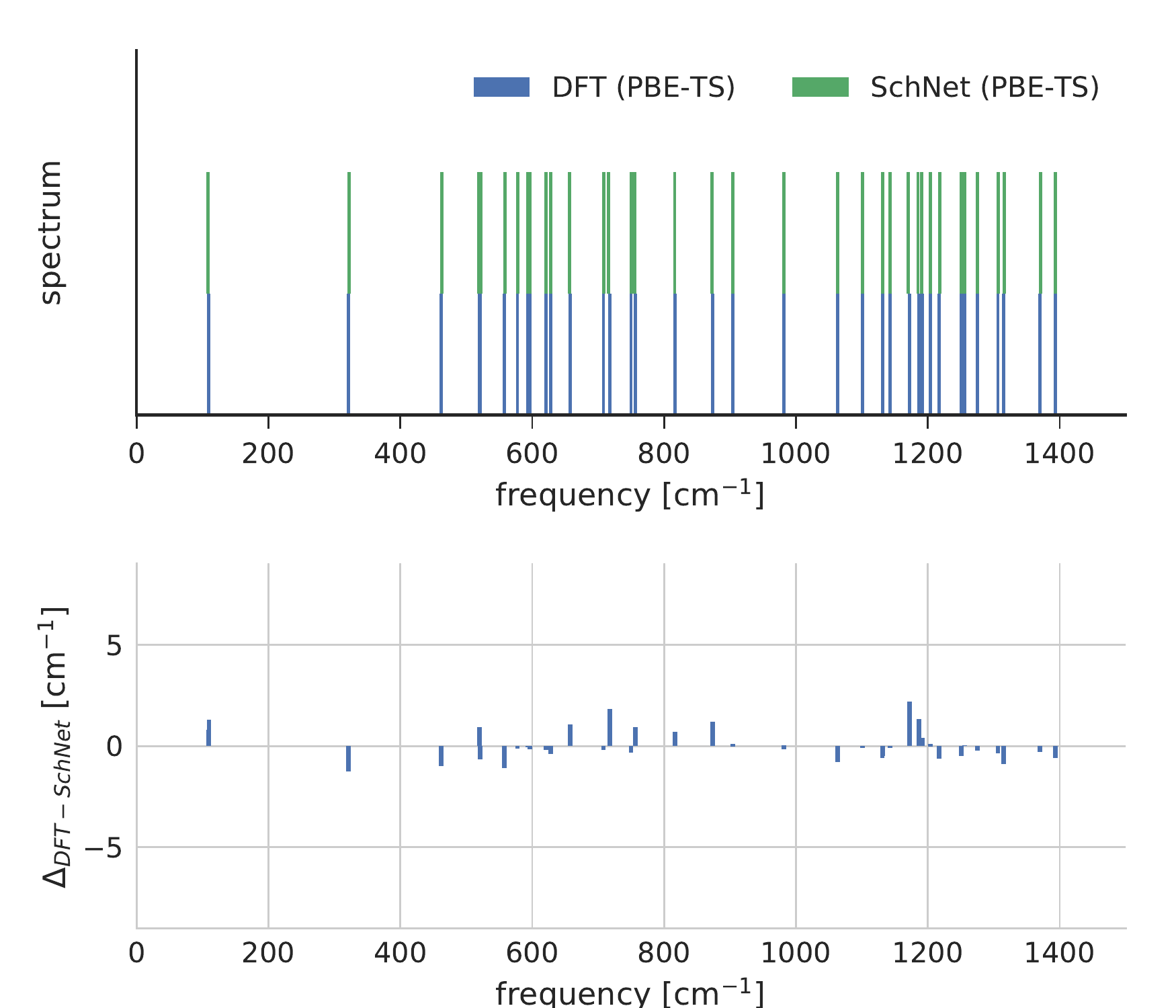}
	\caption{Normal mode analysis of the fullerene C$_{20}$ dynamics comparing SchNet and DFT results.	\label{fig:c20mda}}
\end{figure}
After demonstrating the accuracy of SchNet on the MD17 benchmark set, we perform a study of a ML-driven MD simulation of C$_{20}$-fullerene. 
This middle-sized molecule has a complex PES that requires to be described with accuracy to reproduce vibrational normal modes and their degeneracies. 
Here, we use SchNet to perform an analysis of some basic properties of the PES of C$_{20}$ when introducing nuclear quantum effects.
\new{The reference data was generated by running classical MD at 500 K using DFT at the generalized gradient approximation (GGA) level of theory with the Perdew-Burke-Ernzerhof (PBE)~\citep{PBE1996} exchange-correlation functional and the Tkatchenko-Scheffler (TS) method~\citep{TS2009} to account for van der Waals interactions. For further details about the simulations can be found in Appendix~\ref{app:MDdetails}.}

%The reference data was generated by running classical MD at 500 K using DFT with the generalized gradient approximation (GGA) level of theory with the non-empirical exchange-correlation functional of Perdew-Burke-Ernzerhof (PBE)~\citep{PBE1996} and the Tkatchenko-Scheffler (TS) method~\citep{TS2009} to account for ubiquitous van der Waals interactions as implemented in the FHI-aims code~\citep{FHIaims2009}.}

By training SchNet on DFT data at the PBE+vdW$^{\rm TS}$ level, we reduce the computation time per single point by three orders of magnitude from 11s using 32 CPU cores to 10ms using one NVIDIA GTX1080.
This allows us to perform long MD simulations with DFT accuracy at low computational cost, making this kind of study feasible.

In order to obtain accurate energy and force predictions, we first perform an extensive model selection on the given reference data.
We use 20k C$_{20}$ references calculations as training set, 4.5k examples for early stopping and report the test error on the remaining data.
Table \ref{tab:c20models} lists the results for various settings of number of interaction blocks $T$, number of feature dimensions $F$ of the atomic representations and the energy-force trade-off $\rho$ of the combined loss function.
First, we select the best hyper-parameters $T$, $F$ of the model given the trade-off $\rho=0.01$ that we established to be a good compromise on MD17 (see the upper part of Table \ref{tab:c20models}).
We find that the configuration of $T=6$ and $F=128$ works best for energies as well as forces.
Given the selected model, we next validate the best choice for the trade-off $\rho$.
Here we find that the best choices for energy and forces vastly diverge:
While we established before that energy predictions benefit from force information (see Table~\ref{tab:mdenergies}), we achieve the best force predictions for C$_{20}$-fullerene when neglecting the energies. 
We still benefit from using the derivative of an energy model as force model, since this still guarantees an energy-conserving force field~\citep{chmiela2017machine}.

For energy predictions, we obtain the best results when using a larger $\rho=0.1$ as this puts more emphasis on the energy loss. \new{Here, we select the force-only model as force field to drive our MD simulation since we are interested in the mechanical properties of the $C_{20}$ fullerene.}
Fig.~\ref{fig:c20mda} shows a comparison of the normal modes obtained from DFT and our model. 
In the bottom panel, we show the accuracy of SchNet with the largest error being $\sim$1\% of the DFT reference frequencies. 
Given these results and the accuracy reported in Table \ref{tab:c20models}, we obtained a model that is successfully reconstructing the PES and its symmetries\footnote{Code and trained models are available at: https://github.com/atomistic-machine-learning/SchNet}.

\begin{figure}
	\includegraphics[width=0.95\columnwidth]{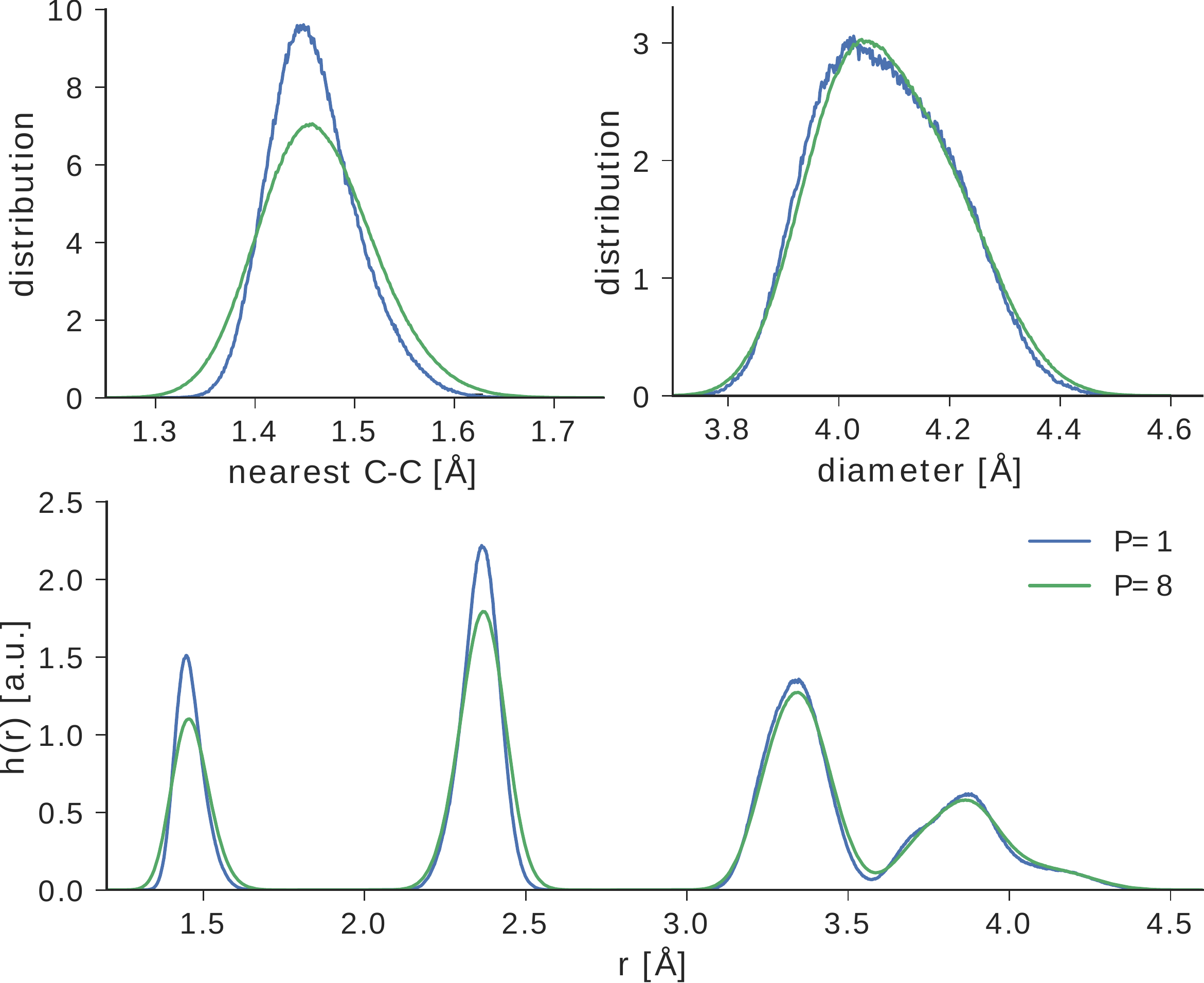}
	\caption{Analysis of the fullerene C$_{20}$ dynamics at 300K using SchNet@DFT. Distribution functions for nearest neighbours, diameter of the fullerene and the atomic-pairs distribution function using classical MD (blue) and PIMD with 8 beads (green).\label{fig:c20mdb}}
\end{figure}
In addition, in Fig.~\ref{fig:c20mdb} we present an analysis of the nearest neighbor (1nn), diameter and radial distribution functions at 300 K for classical MD (blue) and PIMD (green) simulations that include nuclear quantum effects. \new{See Appendix~\ref{app:MDdetails} for further details on the simulation. From Fig.~\ref{fig:c20mdb} (and Fig.~\ref{fig:convBeads}), it looks like nuclear delocalization does not play a significant role in the peaks of the pair distribution function $h(r)$ for C$_{20}$ at room temperature. The nuclear quantum effects increase the 1nn distances by less than 0.5\% but the delocalization of the bond lengths is considerable. This result agrees with previously reported PIMD simulations of graphene~\citep{Igor2017}. However, here we have a non-symmetric distributions due to the finite size of C$_{20}$.}

Overall, with SchNet we could carry out 1.25 ns of PIMD, reducing the runtime compared to DFT by 3-4 orders of magnitude: from about 7 years to \new{less than 7 hours with much less computational resources. Such long time MD simulations are required for detailed studies of mechanical and thermodynamical properties as a function of the temperature, especially in the low temperature regime where the nuclear quantum effects become extremely important}. Clearly, this application evinces the need for fast and accurate machine learning model such as SchNet to explore the different nature of chemical interactions and quantum behavior to better understand molecules and materials.

\section{Conclusions}
\label{sec:conclusions}
Instead of having to painstakingly design mechanistic force fields or machine learning descriptors, deep learning allows to learn a representation from first principles that adapts to the task and scale at hand, from property prediction across chemical compound space to force fields in the configurational space of single molecules.
The design challenge here has been shifted to modelling quantum interactions by choosing a suitable neural network architecture.
This gives rise to the possibility to encode known quantum-chemical constraints and symmetries within the model without loosing the flexibility of a neural network.
This is crucial in order to be able to accurately represent, e.g., the full potential-energy surface and in particular its anharmonic behavior.

We have presented SchNet as such a versatile deep learning architecture for quantum chemistry and a valuable tool in a variety of applications ranging from the property prediction for diverse datasets of molecules and materials to the highly accurate prediction of potential energy surfaces and energy-conserving force fields.
As a variant of DTNNs, SchNet follows rotational, translational and permutational invariances by design and, beyond that, is able to directly model periodic boundary conditions.
Not only does SchNet yield fast and accurate predictions, it also allows to examine the learned representation using local chemical potentials~\citep{schutt2017quantum}.
Beyond that, we have analyzed the atomic embeddings learned by SchNet and found that fundamental chemical knowledge had been recovered purely from a dataset of bulk crystals and formation energies.
Most importantly, we have performed an exemplary path-integral molecular dynamics study of the fullerene C$_{20}$ at the PBE+vdW$^{\rm TS}$ level of theory that would not have been computational feasible with common DFT approaches. 
These encouraging results will guide future work such as studies of larger molecules and periodic systems as well as further developments towards interpretable deep learning architectures to assist chemistry research.

\begin{acknowledgments}
This work was supported by the Federal Ministry of Education and Research (BMBF) for the Berlin Big Data Center BBDC (01IS14013A). Additional support was provided by the DFG (MU 987/20-1), from the European Union's Horizon 2020 research and innovation program under the Marie Sklodowska-Curie grant agreement NO 657679, the BK21 program funded by Korean National Research Foundation grant (No. 2012-005741) and the Institute for Information \& Communications Technology Promotion (IITP) grant funded by the Korea government (no. 2017-0-00451). A.T. acknowledges support from the European Research Council (ERC-CoG grant BeStMo). Correspondence to KTS, AT and KRM.
\end{acknowledgments}

\appendix

\section{\new{Error distributions}}\label{app:distribution}
\begin{figure*}
	\includegraphics[width=\textwidth]{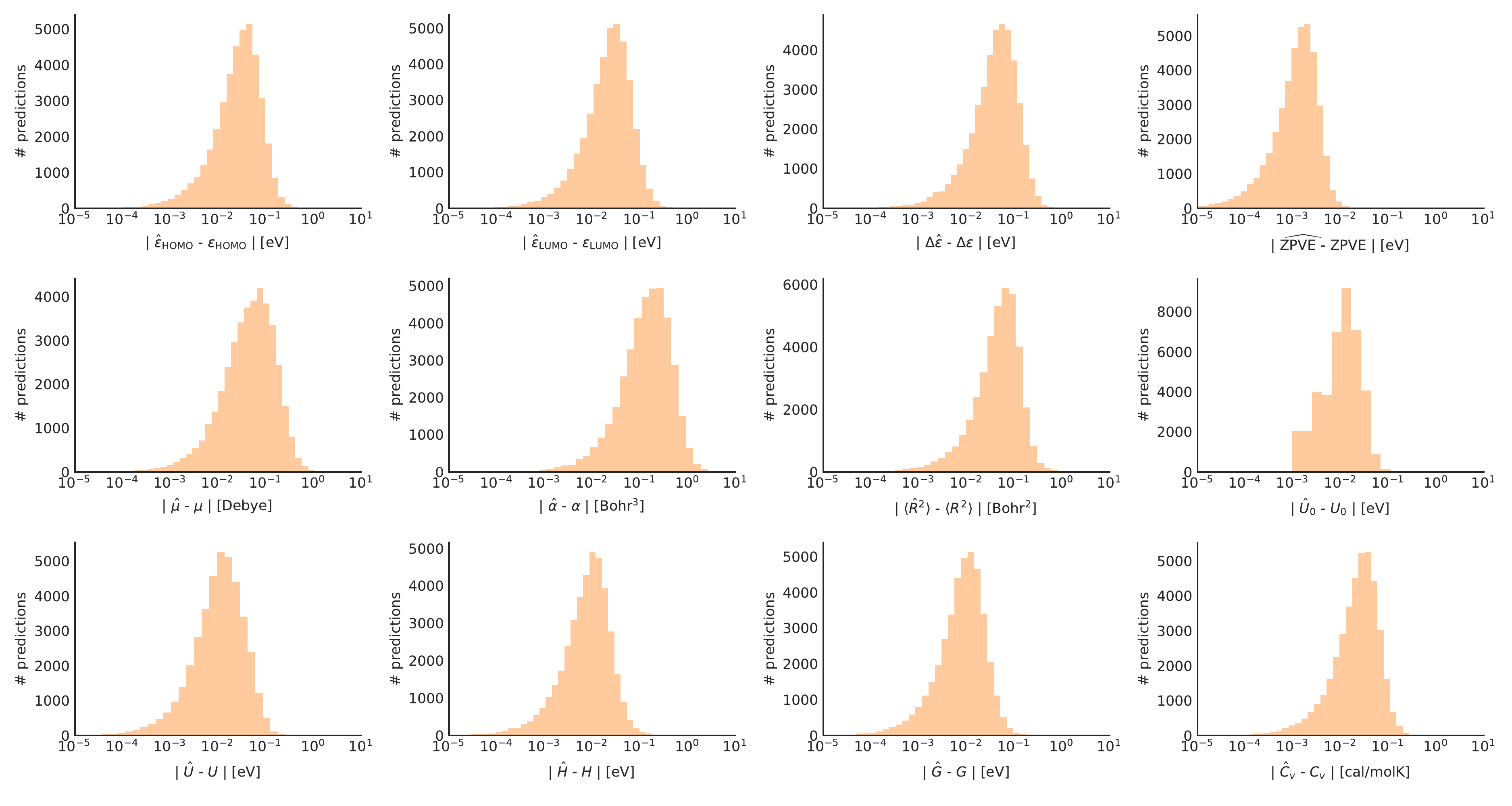}
	\caption{Histograms of absolute errors for all predicted properties of QM9. The histograms are plotted on a logarithmic scale to visualize the tails of the distribution. \label{fig:qm9histogram}}
\end{figure*}
\begin{figure}
	\includegraphics[width=\columnwidth]{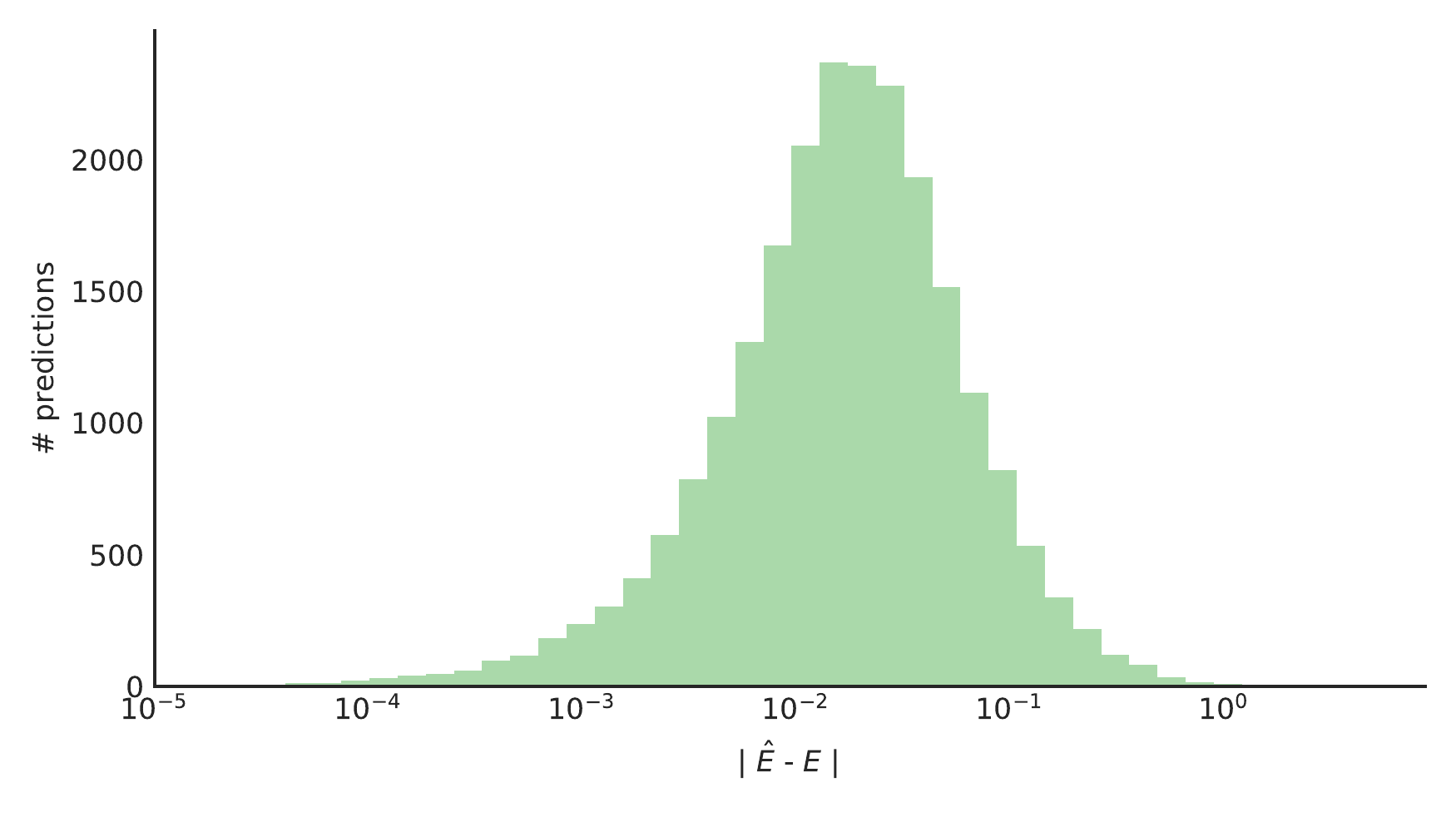}
	\caption{Histogram of absolute errors for the predictions of formation energies / atom for the Materials Project dataset. The histogram is plotted on a logarithmic scale to visualize the tails of the distribution. \label{fig:mphistogram}}
\end{figure}
In Figures \ref{fig:qm9histogram} and \ref{fig:mphistogram}, we show histograms of the predicted properties of the QM9 and Materials Project dataset, respectively.
The histograms include all test errors made across all three repetitions.

\section{\new{MD simulation details}}\label{app:MDdetails}
The reference data for $C_{20}$ was generated using classical molecular dynamics in the NVT ensemble at 500 K using the Nose-Hoover thermostat with a time step of 1 fs. The forces and energies were computed using DFT with the generalized gradient approximation (GGA) level of theory with the non-empirical exchange-correlation functional of Perdew-Burke-Ernzerhof (PBE)~\citep{PBE1996}  and the Tkatchenko-Scheffler (TS) method~\citep{TS2009} to account for ubiquitous van der Waals interactions. The calculations were done using all-electrons with a light basis set implemented in the FHI-aims code~\citep{FHIaims2009}.

\begin{figure*}
	\includegraphics[width=0.8\textwidth]{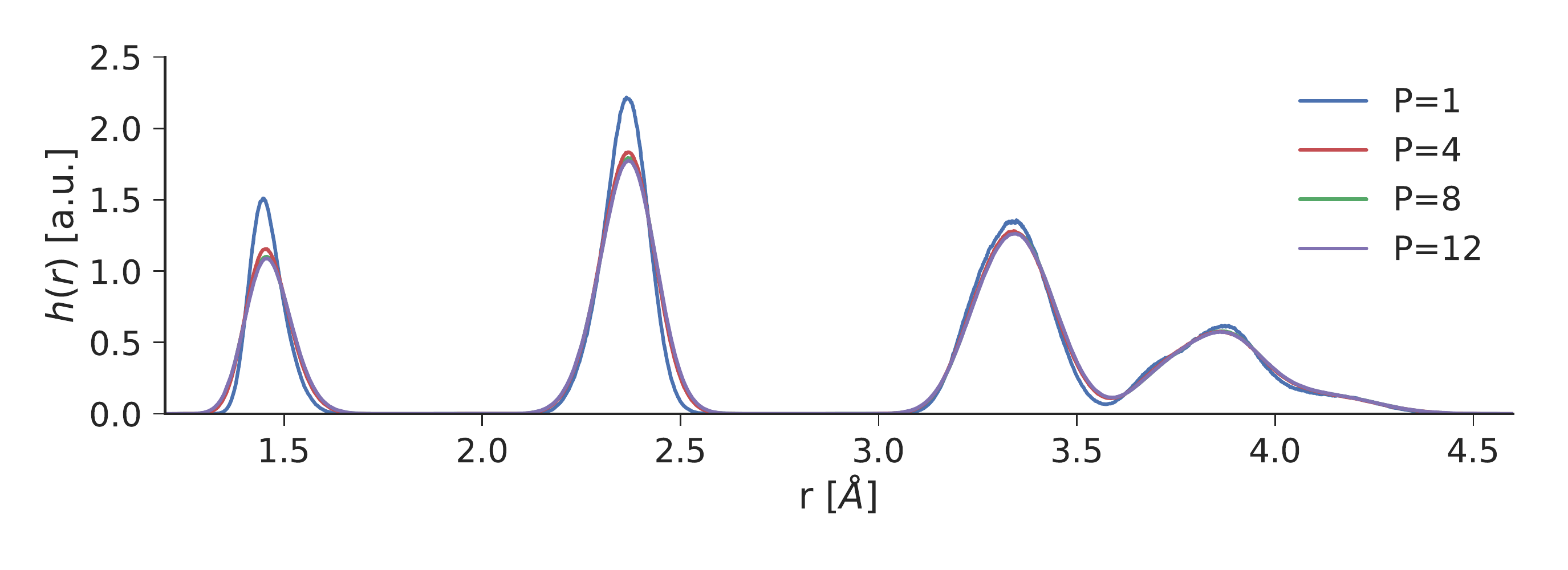}
	\caption{Histograms of absolute errors for all predicted properties of QM9. The histograms are plotted on a logarithmic scale to visualize the tails of the distribution. \label{fig:convBeads}}
\end{figure*}

The quantum nuclear effects are introduced using path-integral molecular dynamics (PIMD) via the Feynman's path integral formalism. The PIMD simulations were done using the SchNet model implementation in the i-PI code~\citep{ceriotti2014pi}. The integration timestep was set to 0.5 fs to ensure energy conservation along the MD using the NVT ensemble with a stochastic path integral Langevin equation (PILE) thermostat\citep{pile2010}. In PIMD the treatment of NQE is controlled by the number of beads, P. In our example for $C_{20}$ fullerene, we can see that at room temperature using 8 beads gives an already converged radial distribution function $h(r)$ as shown in Figure~\ref{fig:convBeads}.

\bibliography{literature}

\end{document}